\documentclass[]{emulateapj}
\usepackage{apjfonts}
\newcommand{\be}{\begin{equation}}
\newcommand{\ee}{\end{equation}}
\newcommand{\bea}{\begin{eqnarray}}
\newcommand{\eea}{\end{eqnarray}}

\newcommand{\Msun}{M_{\odot}}

\shortauthors{CONROY, WHITE, AND GUNN}
\shorttitle{COMPARING GALAXY EVOLUTION MODELS TO OBSERVATIONS}
\begin{document}
\journalinfo{The Astrophysical Journal}
\submitted{Submitted to the Astrophysical Journal}

\title{The propagation of uncertainties in stellar population
  synthesis modeling II: The challenge of comparing galaxy evolution
  models to observations}

\author{Charlie Conroy$^1$, Martin White$^2$, James E. Gunn$^1$}
\affil{$^{1}$Department of Astrophysical Sciences, Princeton
  University, Princeton, NJ 08544, USA \\ $^{2}$Departments of Physics
  and Astronomy, University of California Berkeley, CA 94720, USA}
\begin{abstract}

  Models for the formation and evolution of galaxies readily predict
  physical properties such as the star formation rates, metal
  enrichment histories, and, increasingly, gas and dust content of
  synthetic galaxies.  Such predictions are frequently compared to the
  spectral energy distributions of observed galaxies via the stellar
  population synthesis (SPS) technique.  Substantial uncertainties in
  SPS exist, and yet their relevance to the task of comparing galaxy
  evolution models to observations has received little attention.  In
  the present work we begin to address this issue by investigating the
  importance of uncertainties in stellar evolution, the initial
  stellar mass function (IMF), and dust and interstellar medium (ISM)
  properties on the translation from models to observations.  We
  demonstrate that these uncertainties translate into substantial
  uncertainties in the ultraviolet, optical, and near-infrared colors
  of synthetic galaxies.  Aspects that carry significant uncertainties
  include the logarithmic slope of the IMF above $1\Msun$, dust
  attenuation law, molecular cloud disruption timescale, clumpiness of
  the ISM, fraction of unobscured starlight, and treatment of advanced
  stages of stellar evolution including blue stragglers, the
  horizontal branch, and the thermally--pulsating asymptotic giant
  branch.  The interpretation of the resulting uncertainties in the
  derived colors is highly non--trivial because many of the
  uncertainties are likely {\it systematic}, and possibly correlated
  with the physical properties of galaxies.  We therefore urge caution
  when comparing models to observations.
  
\end{abstract}



\section{Introduction}
\label{s:intro}

Models for the formation and evolution of galaxies have become
dramatically more sophisticated over the past decade.  These models
span a variety of techniques, including the semi--analytic approach,
which couples dark matter halo merger trees to analytic recipes for
the evolution of the baryons \citep[e.g.][]{White91, Kauffmann99a},
and cosmologically--embedded hydrodynamic simulations, which
numerically follow the evolution of dark matter and baryons
self--consistently, with numerical recipes encompassing unresolved
physical processes such as star formation, black hole growth, stellar
and black hole feedback, etc \citep[e.g.][]{Cen92, Katz96, Keres05}.

The primary predictions of these models include quantities such as the
total mass in stars, star formation and metal enrichment histories,
and, in some cases, information on cold gas properties.
Unfortunately, these quantities are not directly available to
extragalactic observers.  Instead, these physical quantities are often
transformed into observable spectral energy distributions via stellar
population synthesis (SPS) techniques \citep[e.g.][]{Tinsley76,
  Bruzual83, Renzini86, Worthey94, Fioc97, Maraston98, Vazdekis99}.
For many applications, SPS thus provides the bridge between models and
observations.

Until now, little attention has been paid to the uncertainties in the
SPS technique in the context of translating galaxy formation model
predictions into observables.  And yet uncertainties in advanced
stages of stellar evolution, stellar spectral libraries, the stellar
initial mass function, and the adopted model for the obscuring effects
of interstellar dust can each be significant components of the total
error budget on the predictions.

Modeling of the attenuation of starlight by dust is a particularly
challenging aspect in the SPS technique.  Such models fall into
roughly two categories.  On the one hand there are approaches that use
physical models for the properties of grains
\citep[e.g.][]{Weingartner01, Zubko04}, coupled to assumptions about
the geometry of the dust with respect to the stars
\citep[e.g.][]{Silva98, Devriendt99, Gordon01, Tuffs04, Jonsson06,
  Galliano08}.  These models rely on radiative transfer calculations
and are therefore computationally expensive.  Their computational cost
has given rise to the second class of models, which are
physically--motivated, but are in essence phenomenological.  Models of
the second class typically adopt an attenuation curve with a
normalization, and, in principle, shape, that depends on the age of
the system \citep[e.g.][]{Charlot00}.  This effective attenuation
curve is then applied uniformly to all stars of a given age.

Common to both approaches is the fact that the required inputs are in
many cases seriously underconstrained.  An accurate dust model
requires knowledge of the grain composition and size distribution, and
the wavelength--dependence of the grain albedo \citep{Draine03}, yet
these quantities are extremely difficult to constrain in the Galaxy
\citep[e.g.][]{Weingartner01}, let alone in external systems.  It is
expected that grain properties will be functions of metallicity and
the local intensity of ultraviolet radiation --- in addition to other
variables--- both of which will vary from galaxy to galaxy.  The
geometry of the dust, including large--scale inhomogeneities and the
proximity of the dust to the stars, is not known at all outside of the
Local Group, and not known with the requisite precision in our own
Galaxy.  Yet the geometry of dust has a substantial effect on the net
attenuation of starlight.  The lifetime of molecular clouds is another
important parameter because it controls the length of time that young
stars are heavily dust obscured, and yet this timescale is not known
to better than an order of magnitude, and will likely depend on
quantities such as metallicity and local star formation rate
\citep{Blitz80, McKee07}.

\citet{Fontanot09} have recently compared a physical dust model that
employs radiative transfer to simple phenomenological prescriptions.
These authors find substantial differences between the analytic
calculations and the more complex physical dust model.  While
comparisons of this type are essential for informing analytic
prescriptions, our belief is that the physical dust models themselves
are sufficiently uncertain to warrant a flexible approach to the
obscuring effects of dust.

Uncertainties in stellar evolution calculations can also impact the
translation from models to observables \citep[e.g.][]{Tinsley80,
  Charlot96a, Charlot96b, LeeHC07, Yi03a, Conroy09a}.  Recently,
\citet{Tonini09} have investigated the impact of the
thermally--pulsating asymptotic giant branch (TP--AGB) phase on the
observational predictions of a semi--analytic galaxy evolution model.
They compare a popular SPS model that does not include the TP--AGB
phase of stellar evolution to the model of \citet{Maraston05} where
this important but uncertain phase is handled with care.  These
authors find that the near--IR colors of model galaxies can differ by
as much as two magnitudes between the two SPS models, and that the
differences increase with redshift.  These results highlight the
importance of carefully accounting for uncertain aspects of SPS when
translating galaxy formation model predictions into observables.

In the present work we extend the spirit of Tonini et al. by
considering a broader array of uncertainties in stellar evolution,
uncertainties in the stellar initial mass function (IMF), and the
substantial uncertainties associated with the treatment of
interstellar dust.  We focus on two synthetic galaxies drawn from a
recent semi--analytic model of galaxy evolution.  These galaxies were
chosen to represent a typical bright star--forming galaxy and a
typical bright passively evolving galaxy, both at $z=0$.  For the
purposes of this work, the main difference between these two types of
galaxies is the presence or absence of young, hot stars and the
average stellar age.

This paper is the second in a series focusing on uncertainties in
stellar population synthesis.  In this series, particular attention is
focused on how those uncertainties effect our knowledge of galaxy
formation and evolution.  In Paper I \citep{Conroy09a} we examined the
impact of SPS uncertainties on deriving physical properties of
observed galaxies, including star formation rates, ages, and stellar
masses.  This paper considers the inverse problem: the uncertainties
in translating model galaxies into observables.

This paper continues with $\S$\ref{s:sps}, where we discuss the
salient ingredients of SPS, including a detailed description of our
dust treatment.  $\S$\ref{s:sam} briefly describes the synthetic
galaxies and the semi--analytic model from which they are derived.
$\S$\ref{s:res} contains our main results, where we investigate the
uncertainties in the predicted UV, optical, and near--IR colors of
synthetic galaxies due to uncertainties in SPS.  Following this
sobering assessment, in $\S$\ref{s:disc} we comment on several
implications.  A summary is provided in $\S$\ref{s:sum}.  All
magnitudes are in the $AB$ system \citep{Oke83}.  Where necessary a
Hubble constant of $H_0 = 100h$ km s$^{-1}$ Mpc$^{-1}$ is assumed.

\section{Stellar Population Synthesis}
\label{s:sps}

Our SPS treatment closely follows that of \citet{Conroy09a}, to which
the reader is referred for details.  In brief, the SPS code uses the
latest stellar evolution tracks from the Padova group
\citep{Marigo07a, Marigo08}, which follow stellar evolution from the
main sequence through the thermally--pulsating asymptotic giant branch
(TP--AGB) phase.  Evolutionary calculations exist for metallicities in
the range $10^{-4}<Z<0.030$, for ages $10^{6.6}<t<10^{10.2} $ yrs, and
for initial masses $0.15\leq M\leq67 \,\Msun$.  The stellar spectral
libraries are primarily those of the empirically--calibrated
theoretical BaSeL3.1 library \citep{Lejeune97, Lejeune98, Westera02},
supplemented with empirical TP--AGB spectra from the library of
\citet{Lancon02}.  The initial stellar mass function (IMF) of
\citet{Kroupa01} is adopted.  Our SPS code is open--source and
publicly available\footnote{\texttt
  www.astro.princeton.edu/$\sim$cconroy/SPS/}.

We seek to assess the importance of uncertainties in SPS in
translating the predictions of galaxy evolution models into
observables.  In our approach we will focus on two distinct classes of
uncertainties.  The first class are uncertainties in the single
stellar populations (SSP) resulting from isochrone synthesis,
including both the stellar evolution tracks and the IMF.  The second
class concerns the treatment of reddening by interstellar dust.

\subsection{SSP uncertainties}

\subsubsection{Stellar evolution (isochrone) uncertainties}

Uncertainties in the isochrones are quantified as in
\citet{Conroy09a}.  The position of TP--AGB stars in the HR diagram is
substantially uncertain, both observationally and theoretically.  We
thus introduce two variables that amount to shifts in ${\rm
  log}(T_{\rm eff})$ and ${\rm log}(L_{\rm bol})$ with respect to the
default stellar evolution tracks: $\Delta_T$ and $\Delta_L$,
respectively.  As discussed in \citet{Conroy09a}, these variables
effectively encompass uncertainties not only in the stellar evolution
tracks but also uncertainties in the associated spectral energy
distributions and circumstellar dust enshrowding these stars.  Since
these stars have relatively low effective temperatures, they
contribute substantially to integrated spectra only at
$\lambda\gtrsim7000$\AA.

Blue straggler (BS) stars and blue horizontal branch (BHB) stars are
almost universally neglected when translating galaxy evolution models
into observables, although their importance has been demonstrated when
modeling observed spectra \citep[see e.g.][]{Jimenez04, Maraston05,
  Li08a, Li09a, Conroy09a}.  We introduce two additional parameters to
quantify these uncertain stellar phases: the specific frequency of
BSs, $S_{\rm BS}$, defined as the number of BSs per unit HB star, and
the fraction of HB stars that are blue, $f_{\rm BHB}$.  In this
context a blue HB star is any HB star extending blueward of the red
clump identified in standard stellar evolution tracks.  In our
treatment this extended blue component is populated uniformly in ${\rm
  log}(T_{\rm eff})$ from the red clump to $T_{\rm eff}=10^4$ K.

We do not allow for any of our parameters to be age--dependent, but
note that one may interpret a given parameter value as being
representative of some effective population age.

\subsubsection{IMF uncertainties}
\label{s:imf}

We will also explore the importance of the logarithmic slope of the
IMF on the derived colors of galaxies.  It is widely understood that
the IMF has a significant impact on the mass--to--light ratio of
galaxies, but it's impact on the SEDs of galaxies is less appreciated
\citep[or rather, has become less appreciated with time, see
e.g.,][]{Tinsley80}.  It is important to consider the influence of the
IMF on the translation between models and observations because of the
difficulty in measuring the IMF in the solar neighborhood, let alone
in external galaxies.

For example, \citet{Kroupa01} compiled a number of observations of the
logarithmic slope of the IMF in the solar neighborhood and found
$\alpha=2.3\pm0.7$ for $m>1\Msun$, at $99\%$ confidence, where the IMF
is parameterized as $\xi(m)\propto m^{-\alpha}$.  It is important to
realize that this uncertainty is only statistical.  As discussed by
\citet{Kroupa01}, corrections due to unseen binary companions will
depend on $\alpha$, and will result in an increase in $\alpha$ of the
order of $0.3$, depending on the details of the binary companions.
Further complications arise near $1\Msun$ because the observational
techniques for determining the IMF at this mass regime rely on star
counts in the field.  At $\sim1\Msun$ the stellar lifetime is of order
the age of the Galactic disk, and so evolutionary corrections, which
are substantial, rely on a detailed knowledge of the star formation
history of the disk.  In addition, because the age of $1\Msun$ stars
is of order a Hubble time, the IMF near this mass matters for
understanding galaxy SEDs.  The uncertainty on $\alpha$ at
$\sim1\Msun$ is so large that data points near $1\Msun$ are often
excluded from fits \citep{Kroupa01}.

The physical conditions of the Universe relevant to star formation
were very different at higher redshift, including higher ISM pressures
\citep{Liu08}, lower metallicities \citep{Erb06c}, and higher CMB
temperatures. The latter will start to effect the fragmentation and
collapse of gas around $z\sim5$ \citep{Larson05}.  Some authors have
suggested that these conditions will result in a top--heavy (or
bottom--light) IMF at earlier times, owing essentially to a larger
Jeans mass \citep[e.g.][]{Larson98, Larson05}.  Recent observational
evidence, including the abundance patterns of metal--poor stars in the
Milky Way \citep{Lucatello05, Tumlinson07a}, evolution in the
fundamental plane \citep{vanDokkum08}, and a comparison between the
cosmic mass density evolution and star formation rate density
evolution beyond $z\sim2$ \citep{Dave08}, is beginning to support this
picture, however tentatively \citep[for an alternative, less exotic
explanation of the last point, see][]{Reddy09}.

Recent observational results in the local Universe has also pointed
toward an IMF that varies with environment.  \citet{Meurer09} recently
demonstrated that the ratio between $H\alpha$ flux and near-UV
luminosity varies systematically with galaxy surface brightness.  This
ratio should be constant for a universal IMF.  The data can be
explained by an IMF slope at high masses that varies over the range
$1.5<\alpha<4.0$, or by a systematically varying upper mass cut--off
over the range $\approx20<M_{\rm up}<100\,\Msun$.

We follow \citet{Kroupa01} in the parameterization and best--fit
logarithmic slopes of the IMF.  In the following sections we will
explore the effects of varying the IMF only for masses $>1\Msun$.
Variation is limited to $>1\Msun$ because the main--sequence turn--off
mass is $1\Msun$ for a 12 Gyr SSP, and so stars of lower mass never
contribute substantially to the integrated colors (unless pathological
IMFs are considered).  In addition, as will be discussed in
$\S$\ref{s:stevimf}, it is primarily galaxies with a range of stellar
populations that are sensitive to IMF variations.  In such galaxies it
is the relative frequencies of stars with mass $>1\Msun$ that
contribute to the integrated light and therefore only IMF variations
at $>1\Msun$ will have impact on the colors.

\vspace{0.5cm}

\subsection{Dust treatment}
\label{s:dust}

We now turn to the treatment of dust in the SPS technique.  Since we
are interested in restframe wavelengths bluer than $\approx2.4\mu m$
(i.e. the $K-$band), we do not model the reprocessing of radiation by
dust; we only attempt to model its obscuring effects.

The light emerging from a galaxy can be modeled as
\noindent
\be
\label{eqn:flux}
F_\lambda(t) = \int_0^t\,\Psi(t-t')S_\lambda(t',Z(t-t'))\, e^{-\tau_\lambda(t')}\,{\rm d}t',
\ee
where $\Psi$ is the SFR, $S_\lambda$ is the SSP spectrum, $F_\lambda$
is the resulting composite spectrum, $\tau_\lambda$ is the optical
depth, and the integration variable is the age of stellar populations.
Note that this formulation allows for metallicity evolution with time,
$Z(t)$.

The obscuring effects of dust are encompassed in the optical depth
$\tau_\lambda(t)$.  A toy model for the effective optical depth was
presented in \citet{Charlot00} and has since become quite
popular\footnote{This is one of the dust models provided in the
  popular \citet{Bruzual03} SPS code, and therefore is a standard dust
  model for many investigators.}.  The physical motivation for their
model is the following.  Stars are born in molecular clouds and thus
the light from young stars is heavily attenuated by dust in the cloud.
At later times the stars will no longer experience attenuation due to
their birth cloud, either because they will have wandered out of the
cloud, or because the cloud will have evaporated.  Stars are thus
subject to attenuation that varies with time --- young stars are
obscured by the dense molecular clouds in which they form in addition
to the diffuse ISM, while the light from older stars is attenuated
only by the diffuse ISM.  Motivated by this physical picture, the
optical depth is parameterized as:
\noindent
\be
\tau_\lambda(t) = \left\{ \begin{array}{lc}
 \tau_1(\lambda/5500 {\textrm \AA})^{-\delta_1} &\, t \leq t_{\rm esc}  \\
 \tau_2(\lambda/5500 {\textrm \AA})^{-\delta_2} &\, t > t_{\rm esc},
\end{array} \right.
\ee
\noindent
where $t_{\rm esc}$ is the timescale over which young stars reside in
their natal clouds.  While there is ample evidence that the extinction
curve slope depends on environment \citep[e.g.][]{Mathis90}, herein we
fix $\delta_1=\delta_2\equiv\delta$, which has become a common
assumption \citep[e.g.][]{Charlot00}\footnote{In future work we will
  investigate this assumption in detail.  For the present work, we are
  interested in quantifying the uncertainties associated with the dust
  model, and so it matters little if we vary $\delta_1$ and $\delta_2$
  separately or in conjunction.}.  \citet{Charlot00} advocate
$\delta=0.7$, ${\rm log}(t_{\rm esc}/{\rm yrs})=7.0$, $\tau_1=1.0$,
and $\tau_2=0.3$, although it is clear from their own figures that
these parameters may plausibly vary over a wide range.  We choose to
leave these parameters free with a range motivated by a variety of
data (see discussion in $\S$\ref{s:prior}).

In reality each of the parameters $\tau_1$, $\tau_2$, $\delta$, and
$t_{\rm esc}$ will take on a range of values for a given galaxy.  For
example, $\tau_1$ will depend on the location of the young star
cluster with respect to the molecular cloud within which it's
embedded.  If the stars are born near the edge of the cloud, as is
often observed within the Galaxy \citep{Israel78}, then $\tau_1$ will
depend strongly on the relative orientation of the observer with
respect to the star cluster---cloud configuration.  A more accurate
dust model would therefore allow for a {\it distribution} in each of
these quantities.  Owing to the non--linear relation between these
parameters and the resulting galaxy spectrum, the use of distributions
rather than a mean values will have important consequences.  These
issues will be considered in detail in future work; in the present
work we consider only mean values for $\tau_1$ and $\delta$.

The optical depth associated with the diffuse ISM, $\tau_2$, will be
handled separately.  Many treatments of dust attenuation adopt the
reasonable assumption that the optical depth of dust is proportional
to the column density of metals \citep[e.g.][]{Guiderdoni87}.  Of the
parameters listed above, $\tau_2$ is therefore the parameter most
readily predicted by galaxy evolution models.  In our main results, we
therefore choose not to marginalize over this important parameter.  It
must be noted, however, that this parameter is not at all easily or
reliably predictable in the majority of galaxy evolution models owing
to the enormous dynamic range required to adequately resolve the
relevant physical processes leading to the production of metals.  And,
even if the quantity and spatial distribution of metals an be
accurately predicted, the relationship between metals and dust is in
reality quite complex owing to the variety of heating and cooling
processes in the ISM relevant for dust formation and destruction.

The attenuation law required in our formulation represents the net
fraction of starlight photons removed by dust, as seen by a distant
observer.  It is different than true extinction, which quantifies the
fraction of photons removed along the line of sight to a single star.
The difference is that extinction is due to both true absorption and
scattering, while attenuation is primarily true absorption and
geometric effects.  In other words, while photons from a single star
may be scattered out of the line of sight to an observer, when viewing
an entire galaxy those scattered photons will on average (depending on
geometry) be seen by a distant observer.  Extinction can be measured
in the Milky Way and Magellanic Clouds, but we require attenuation
when applying dust corrections to external galaxies, and therefore
require accurate knowledge of the net scattering effects of dust.  See
\citet{Calzetti01} for detailed discussion of these and related
issues.

It is well--known that the Milky Way and Magellanic Cloud extinction
curves are approximately a one parameter family \citep{Cardelli89,
  Fitzpatrick99}.  In the optical and near--IR these curves are
well--characterized by a power--law.  The net attenuation in local
starburst galaxies is also approximately a power--law from the far--UV
to $\approx1\mu m$ \citep{Calzetti94, Calzetti00}.  Motivated by the
qualitative aspects of these observational results, we have adopted a
power--law attenuation law in our dust model.  For the purposes of
this paper it is not essential that our parameterization provide the
most accurate reflection of the underlying attenuation curve; rather,
it is important only that the parameterization be flexible enough to
encompass the uncertainties, or variations, in the attenuation law.
Or, put differently, for our purposes we are primarily interested in
an accurate parameterization of the variance, and not the mean, of the
attenuation law.

There is an additional feature in the Galactic and Magellanic Cloud
extinction curves that deserves mention.  The strong, broad absorption
feature at $2175$\AA\, is ubiquitous in the Galactic and LMC
extinction curves, and has even been detected at $z\approx2$
\citep{Noll05, Ardis09, Noll09}.  It is however conspicuously absent
in the attenuation curves within local starbursts \citep{Calzetti94}
and in the extinction curves through most sightlines of the SMC
\citep{Pei92}.  It is thought that this feature is due to polycyclic
aromatic hydrocarbons \citep{Draine03}, and that it may vary with
local environment and star formation rate \citep{Gordon03}.

This feature happens to lie within the {\it GALEX} NUV band for
low--redshift galaxies.  Its presence or absence can change the
attenuation in the NUV band by a magnitude or more.  We do not
consider the uncertainties associated with this feature herein, but
simply note that including it would likely double the resulting
uncertainties in the NUV band.  Future work will consider this issue
in further detail.

\subsubsection{Large--scale dust distribution}
\label{s:clumps}

The prescription described above assumes that stellar populations are
obscured by a uniform screen of dust.  In reality, one expects the
large--scale diffuse ISM to exhibit considerable variation.  This
effect is important because a clumpy ISM will result in a lower
effective optical depth compared to a uniform screen, for the same
total amount of gas and dust.  Moreover, the effect of clumpiness is
wavelength--dependent, implying that the clumpiness of the ISM cannot
be treated as a simple modulation of $\tau_2$ \citep{Natta84, Witt96}.
An accurate accounting of the large--scale inhomogeneities of the ISM
is therefore a crucial element of any dust model
\citep[e.g.][]{Bruzual88, Witt92, Varosi99, Witt00, Gordon01,
  Jonsson06}.  We now describe our treatment of a clumpy ISM.

We define $n$ as the density enhancement of the ISM along a given
sightline, relative to a uniform screen.  Then Equation \ref{eqn:flux}
becomes:
\noindent
\be
F_\lambda(t) =  \int_{0}^t \int_0^\infty P(n)e^{-n\tau_{\lambda}(t')}\,\Psi(t-t')S_\lambda(t',Z(t-t'))\,{\rm d}n\,{\rm d}t',
\ee
\noindent
where $P(n)$ is the probability that an SSP is behind a column density
of dust enhanced by a factor $n$ relative to the uniform screen.  The
above relation holds as long as the optical depth is proportional to
the column density of dust.

We do not modify the treatment of obscuration around young stars where
$t<t_{\rm esc}$.  Therefore:
\noindent
\be
P(n) \equiv \left\{ \begin{array}{ll}
 \delta(n-1)  &\, t \leq t_{\rm esc} \,\,{\rm yr} \\
 P_2  &\, t > t_{\rm esc} \,\,{\rm yr},
\end{array} \right.
\ee
\noindent
where $\delta(x)$ is the Dirac delta function.

There is a constraint that must be satisfied for $P_2$:
\noindent
\be
\label{eqn:c1}
\int n\,P_2\,{\rm d}n  =  1.
\ee
\noindent
This equation ensures that the clumping of dust does not
alter the total amount of dust in the galaxy.

The parameter $\tau_2$ takes on a somewhat different meaning in the
context of a clumpy ISM.  Formally, it describes the attenuation that
would be experienced if the ISM were smoothed to a uniform density.
Physically, $\tau_2$ can be thought of as simply controlling the total
dust mass in the galaxy --- e.g., keeping $\tau_2$ fixed while varying
$P_2$ implies that the total amount of dust is held constant, and only
its spatial distribution is changed.

There are very few direct observational constraints on the column
density distribution of gas and dust in the Galaxy, let alone in
external systems.  \citet{Berkhuijsen08} measured the PDF of diffuse
atomic gas in the solar neighborhood.  They found that the
distribution of densities is consistent with lognormal with a
dispersion of $\sim0.3$.  A lognormal distribution of column densities
naturally arises from simulations of turbulence in an isothermal ISM
\citep{Vazquez-Semadeni94, Scalo98, Nordlund99, Ostriker01}, with a
width that scales with the Mach number.  \citet{Fischera03} attempt to
fit the observed attenuation law derived by \citet{Calzetti00} with a
model that includes a clumpy ISM.  Their best--fit values for the
dispersion range from $0.6<\sigma<2.2$.

Motivated by these results, we adopt a lognormal PDF for the
distribution of column densities (i.e. $P_2$ in Equation \ref{eqn:c1}
is a Gaussian in the variable ${\rm log}(n)$).  Equation \ref{eqn:c1}
implies a constraint on the relation between the mean and variance of
a lognormal PDF that we fit with a fifth--order polynomial:
\noindent
\bea
\label{eqn:sigmaclump}
 \mu &\approx& 0.0709\,\sigma_{\rm clump}-1.68\,\sigma_{\rm clump}^2+1.56\,\sigma_{\rm clump}^3-{}  \nonumber\\
& & {}\,\,1.96\,\sigma_{\rm clump}^4+0.886\,\sigma_{\rm clump}^5,
\eea
\noindent
where $\mu\equiv\mu_{{\rm log}n}$ is the logarithmic mean and
$\sigma^2_{\rm clump}$ is the logarithmic variance.  This fitting
function is valid over the range $0.0\leq\sigma_{\rm clump}<1.3$.
Note that our default assumption of a uniform distribution of dust is
recovered in the limit where $\sigma_{\rm clump}=0$ (i.e. in this
limit $P_2=\delta(n-1)$).  The dispersion characterizing the lognormal
PDF is therefore the single variable controlling the clumpiness of the
ISM in our model.

In reality one might expect the distribution of column densities to be
more complex than lognormal.  In the limit that dust lies at the
mid--plane of a galactic disk \citep[e.g.][]{Dalcanton04}, fully one
half of the disk stars will experience no attenuation.  For galaxies
that contain both a stellar bulge and disk, the majority of the bulge
stars will also experience little or no attenuation, again under the
assumption that most of the dust lies at the mid--plane of the disk.
While we do not include these more realistic geometries in our general
analysis, we will demonstrate in $\S$\ref{s:res1} that they can have a
very important effect on the derived colors of synthetic galaxies.

\begin{figure}[!t]
\plotone{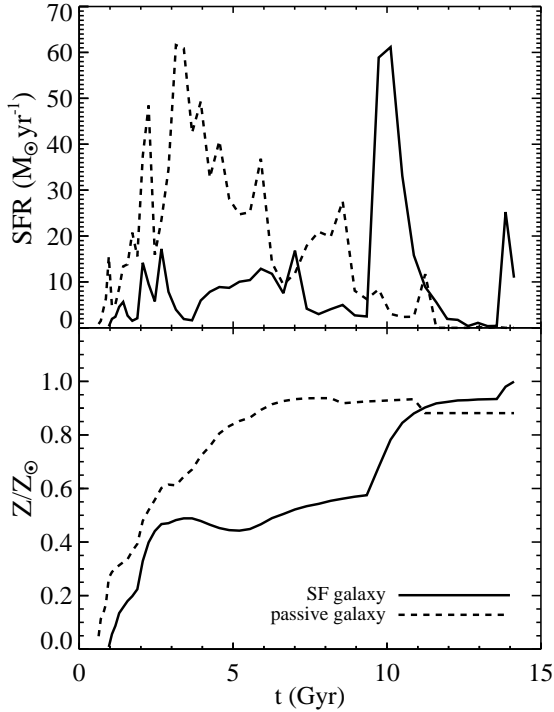}
\vspace{0.5cm}
\caption{Star formation and metal enrichment histories for the two
  synthetic galaxies considered herein: a passive and a star--forming
  (SF) galaxy.}
\label{fig:sfhz}
\vspace{0.5cm}
\end{figure}

\subsection{Parameter ranges and priors}
\label{s:prior}

This section describes and motivates the adopted priors on each
parameter.  In all cases we will assume, for simplicity, that the
prior distribution is flat between the minimum and maximum of the
range.

\begin{figure*}[!t]
\center
\includegraphics[angle=90, width=0.9\textwidth]{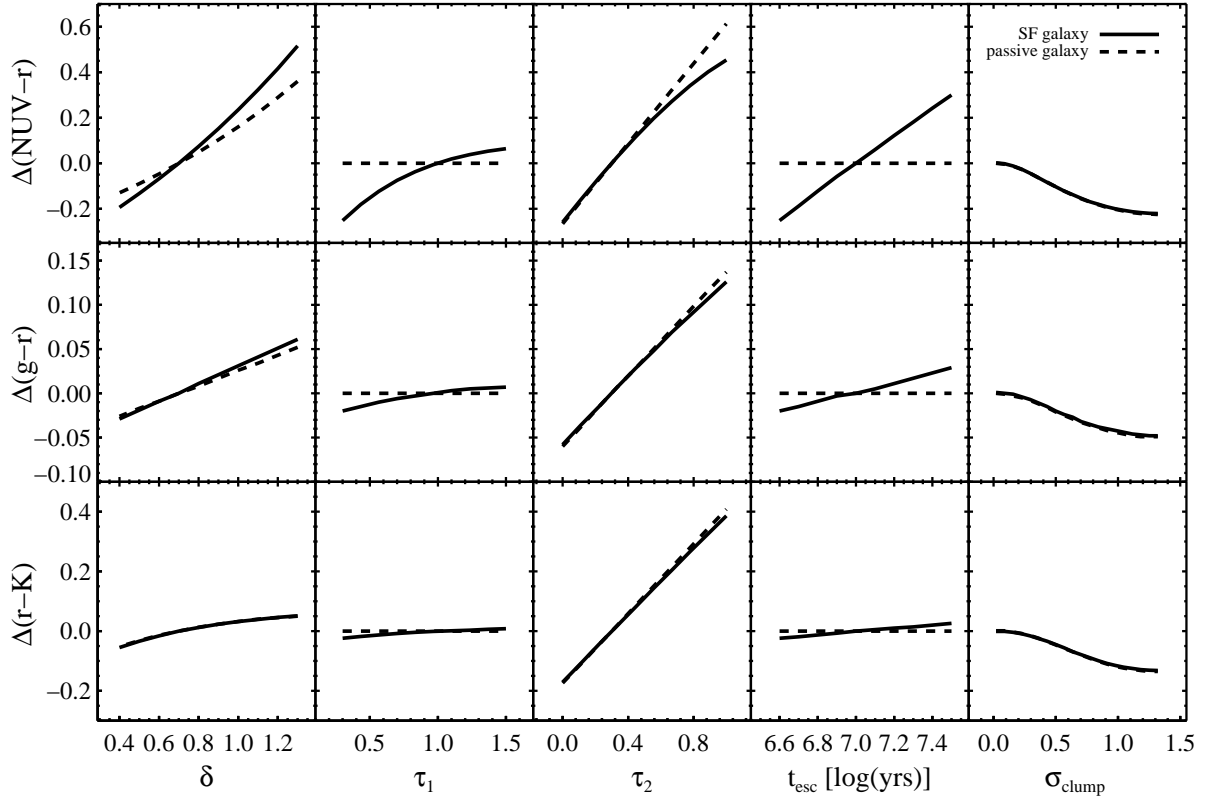}
\vspace{0.5cm}
\caption{Effect of dust parameters on UV, optical, and near--IR colors
  of a star--forming galaxy ({\it solid lines}) and a passive galaxy
  ({\it dashed lines}).  The parameters are described in
  $\S$\ref{s:prior}.  The colors are shown as differences from a
  default model where $\delta=0.7$, $\tau_1=1.0$, $\tau_2=0.3$,
  $t_{\rm esc}=7.0$, and $\sigma_{\rm clump}=0.0$.  In each panel,
  only one parameter is varied.  For the passive galaxy, the colors
  are independent of the parameters $\tau_1$ and $t_{\rm esc}$ because
  there are no young stars in this galaxy and thus these parameters,
  which only effect dust around young stars, are not relevant.  The
  dependence of the colors on $\sigma_{\rm clump}$ is the same for
  both galaxies.}
\label{fig:dust}
\vspace{1.0cm}
\end{figure*}

\begin{itemize}

\item {\bf $\Delta_L$: } Shift in ${\rm log}(L_{\rm bol})$ with
  respect to the default evolutionary tracks of TP--AGB stars.  There
  is a lack of data on the TP--AGB phase outside of the Galaxy and the
  Magellanic Clouds, making it very difficult to even specify a
  reasonable prior range for this parameter.  We adopt the following
  optimistic estimate: $-0.2<\Delta_L<0.2$, and refer the reader to
  \citet{Conroy09a} for motivation of this range.

\item {\bf $\Delta_T$: } Shift in ${\rm log}(T_{\rm eff})$ with
  respect to the default evolutionary tracks of TP--AGB stars.  As
  above, observational constraints are lacking for this parameter.  We
  adopt the optimistic range: $-0.1<\Delta_T<0.1$, and refer the
  reader to \citet{Conroy09a} for details.

\item {\bf $f_{\rm BHB}$: } Fraction of HB stars that are blueward of
  the red clump.  It is well known that metal--poor globular clusters
  ([Fe/H]$\lesssim-1.4$) have extended HB morphologies
  \citep{Harris96}.  In addition, there are several less metal--poor
  globulars ([Fe/H]$\approx-0.6$) with blue horizontal branches
  \citep{Rich97}, and the most metal--rich Galactic star cluster
  known, NGC 6791, has $f_{\rm BHB}\approx0.3$ \citep{Kalirai07}.  In
  our Galaxy, several percent of stars are metal--poor
  \citep{Zoccali03, Zoccali08, Schoerck08}, while in elliptical
  galaxies the fraction may range from $1-10$\% \citep{Worthey96,
    Maraston00}.  \citet{Dorman95} demonstrated that the observed UV
  upturn in elliptical galaxies can be explained by $5-20$\% blue
  (extreme) horizontal branch fractions.  Based on these results, we
  adopt $0.0<f_{\rm BHB}<0.2$.

\item {\bf $S_{\rm BS}$: } Specific frequency of blue straggler stars,
  defined as the number of blue stragglers per unit horizontal branch
  star.  Typical values for $S_{\rm BS}$ range from $0.1-1.0$ for
  globular clusters \citet{Piotto04}.  While much more challenging to
  measure, the frequency of blue stragglers in the field may be as
  high as five \citep{Preston00}.  We adopt the following range for
  $S_{\rm BS}$, which will be high if blue stragglers only exist in
  globular clusters, but appropriate if they originate from binary
  star systems, which are ubiquitous in the field: $0<S_{\rm BS}<2$.

\item {\bf $\alpha_{\rm IMF}$: } Logarithmic slope of the IMF above
  $1\Msun$.  This parameter is not included in our total uncertainty
  budget, but is included in Figure 5 to provide intuition for it's
  importance.  \citet{Kroupa01} quote $\alpha_{\rm IMF}=2.3\pm0.7$ for
  $M>1\Msun$.  The quoted uncertainties are $3\sigma$,
  statistical. However, as pointed out by Kroupa, the corrections to
  $\alpha$ due to unseen binary companions depends on $\alpha$ and is
  of the order of $0.3-0.4$.  Moreover, direct constraints on the IMF
  come {\it only} from the Milky Way and Magellanic Clouds, while we
  require the IMF for star clusters of all ages and metallicities.  In
  addition, there is increasing indirect evidence that the IMF was
  substantially different when the Universe was younger and in
  different environments at low redshift \citep[for a summary, see
  $\S$\ref{s:imf}, and][]{Elmegreen09}.  For these reasons we adopt
  the following range: $1.6<\alpha_{\rm IMF}<3.0$.

\item {\bf $\delta$: } Power--law index of attenuation curve.  In a
  comparison to properties of local starbursts, \citet{Charlot00} find
  that the observations can be well--fit with $0.4<\delta<1.3$.  The
  upper limit also corresponds to the Milky Way extinction curve in
  the optical and near-IR \citep{Fitzpatrick99}, while the lower limit
  roughly approximates the optical region of an extinction curve with
  $R\equiv A_V/E(B-V)\approx6$.  Extinction curves with such high
  values of $R$ are seen along dense sightlines in the Galaxy
  \citep[e.g.,][]{Cardelli89}.  The attenuation law for local
  starbursts can be characterized by a power--law index of
  $\approx0.9$ from the UV to $\approx1\mu m$ \citep{Calzetti00}.
  Motivated by these considerations, we adopt the following range:
  $0.4<\delta<1.3$.

\item {\bf $t_{\rm esc}$: } For star clusters, the transition
  timescale from birth clouds to the diffuse ISM.  This transition
  occurs either because the birth cloud is destroyed
  (i.e. photoevaporated by O--type stars), or because the stars wander
  out of the cloud.  The relative importance of these two processes is
  uncertain but will depend on the mass of the cloud, in addition to
  the local star formation rate.  A variety of observational estimates
  both in the Galaxy and the Local Group suggest that this parameter
  lies in the range $6.5<{\rm log}(t_{\rm esc}/{\rm yrs})<7.5$
  \citep{Israel78, Williams97, Blitz80, Blitz07, McKee07}.  This range
  will be adopted herein.

\item {\bf $\tau_1$: } Optical depth around young stars (where $t\leq
  t_{\rm esc}$).  \citet{Charlot00} found that a variety of data from
  starburst galaxies are bracketed by $0\lesssim\tau_1\lesssim2$.
  \citet{Humphreys78} tabulated $V-$band optical depths for hundreds
  of supergiants and O--type stars in the Galaxy and found a rough
  range of $0.4\lesssim\tau_1\lesssim3$.  Observations of
  extragalactic HII regions find similar results \citep{Israel80,
    vanderhulst88}.  We therefore adopt a range of $0.3<\tau_1<1.5$ as
  the range for this parameter.

\item {\bf $\tau_2$: } Optical depth around old stars (where $t>t_{\rm
    esc}$).  In the context of our dust model, this parameter may be
  considered as a proxy for the dust mass in a galaxy.  In our total
  uncertainty budget, this parameter is not allowed to vary because,
  as discussed in $\S$\ref{s:dust}, it may, with some degree of faith,
  be predicted in galaxy formation models from the surface--averaged
  column density of metals.  We vary this parameter in Figure
  \ref{fig:dust} only to provide intuition for its importance.

\item {\bf $\sigma_{\rm clump}$: } Dispersion of the lognormal PDF
  characterizing the distribution of column densities of the ISM.
  Owing to the absence of constraining data beyond the solar
  neighborhood, we allow the $\sigma_{\rm clump}$ to vary between
  $0.0<\sigma_{\rm clump}<1.3$.  The upper limit is set by the
  practical consideration that the effect of $\sigma_{\rm clump}$ on
  colors tends to saturate near this value, and so allowing larger
  values has no effect on observables (see $\S$\ref{s:res} for
  details).  A value of $0.0$ corresponds to a uniform screen of dust.
 
\end{itemize}

\section{Synthetic Galaxies}
\label{s:sam}

\begin{figure}[!t]
\plotone{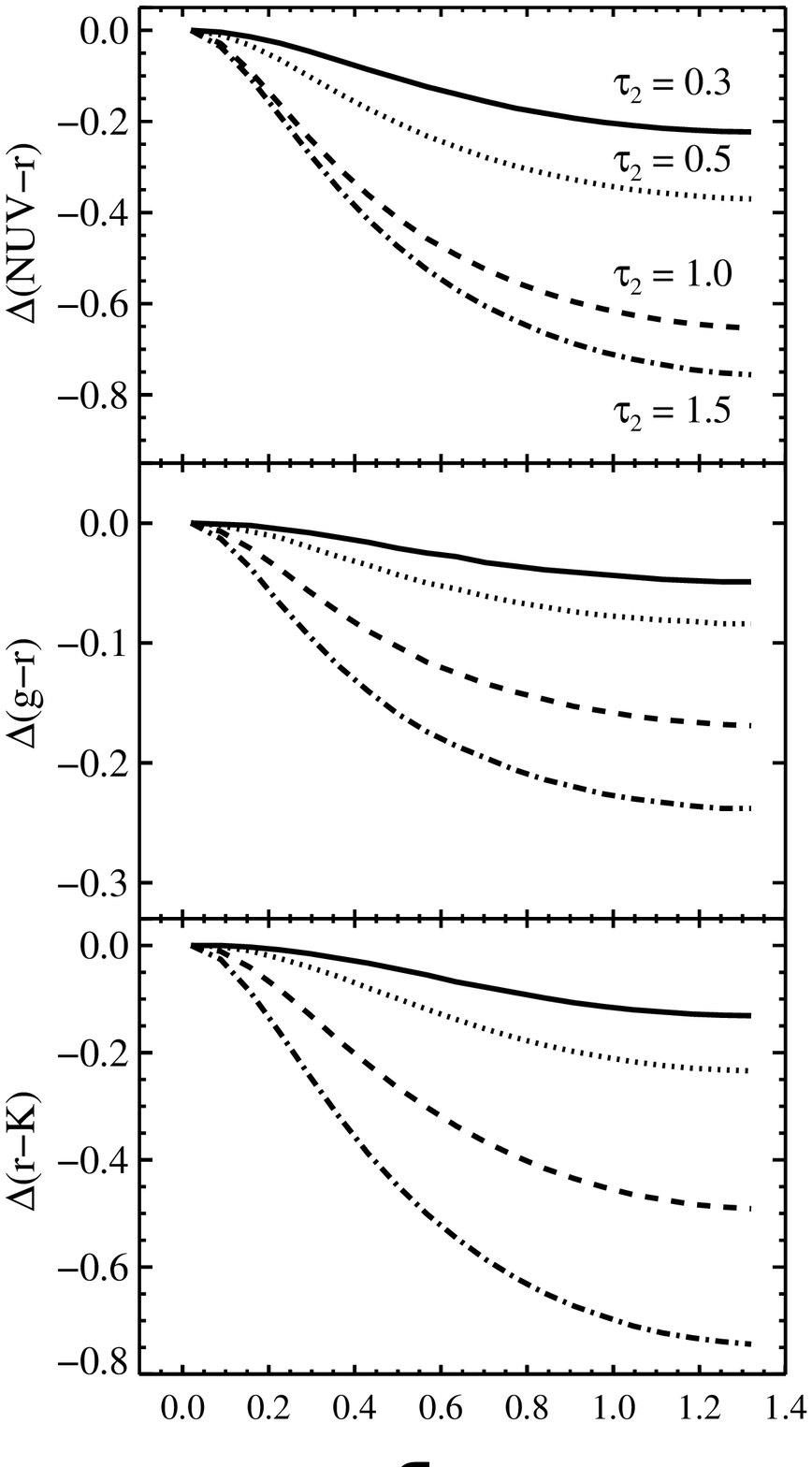}
\vspace{0.5cm}
\caption{Dependence of colors on the dispersion in column densities of
  the ISM (where $\sigma_{\rm clump}=0.0$ corresponds to a uniform
  screen of dust).  Each relation is normalized to the colors at
  $\sigma_{\rm clump}=0.0$.  The relations are shown for several
  values of the optical depth parameter $\tau_2$, as labeled in the
  figure.  The $\tau_2=0.3$ case is identical to the far--right column
  of Figure \ref{fig:dust}.  These results are shown for the
  star--forming galaxy only.  This figure demonstrates that the
  importance of carefully handling the density fluctuations in the ISM
  increases dramatically at higher mean optical depths, i.e. for
  larger total dust masses.}
\label{fig:clumps}
\vspace{0.2cm}
\end{figure}

As described in the Introduction, our goal is to investigate to what
extent uncertainties in the SPS technique hamper the translation of
synthetic galaxies into observables.  The outputs of galaxy evolution
models typically include star formation and metallicity--enrichment
histories of galaxies, and, increasingly, information about the
spatial distribution of gas as well.

We make use of the Millennium Simulation
database\footnote{http://www.mpa-garching.mpg.de/millennium/} to query
the outputs from the semi--analytic galaxy formation model of
\citet{DeLucia07}.  This model couples analytic recipes for gas
cooling, star formation, supernova feedback, AGN heating, galaxy
merging and metal enrichment, to a large, cosmological,
dissipationless $N$--body simulation \citep[the `Millennium
Simulation';][]{Springel05}.  The details of their model do not concern
us here --- we are merely interested in extracting a reasonable star
formation and metal--enrichment history for a typical passive and
star--forming galaxy.

Note that spatial information on the scale of a galaxy is not
available for the galaxy evolution model we consider, in particular
the spatial distribution of the gas with respect to the stars is not a
prediction of the model.  This is in contrast to modern cosmological
hydrodynamic simulations.  Of course, without this detailed spatial
information it is extremely difficult to accurately account for the
attenuation of starlight by dust.

For the bulk of our analysis we will focus on two galaxies that
represent both passive (non--star forming) and star forming galaxies.
The star formation and metal--enrichment histories for these two
galaxies are shown in Figure \ref{fig:sfhz}.  Notice that the passive
galaxy experienced the bulk of its star formation in the distant past
(and has experienced no star formation in the past $\approx 4$ Gyr),
while the star--forming galaxy has experienced multiple episodes of
intense star formation super--imposed on a relatively quiescent level
of star formation.  Assuming a \citet{Kroupa01} IMF\footnote{Given a
  star formation history, the IMF effects the total stellar mass only
  insofar as it controls the fraction of initial stellar mass formed
  that is lost due to supernova and stellar winds, via the relative
  proportions of low and high mass stars.}, the passive
(star--forming) galaxy has a stellar mass at $z=0$ of ${\rm
  log}(M/\Msun)=11.1 (10.9)$.

When considering population statistics (i.e luminosity functions; LFs)
we make use of a subsample of the full galaxy formation model.  We
draw galaxies at random from the full model with a sampling rate of
$10^{-4}$, resulting in $\approx7000$ synthetic galaxies.  This
particular number was chosen as a compromise between adequately
sampling the underlying distribution and computational cost.  Since
the models do not predict dust content, we must specify a recipe for
adding dust to the galaxy formation model.  We make the simple
assumption that $\tau_2=\sqrt{SFR}/2$, where SFR is measured in
$M_\Sol$/yr, and then assume that $\tau_1=\tau_2/0.3$
\citep{Charlot00, Kong04}.  These simple assumptions do not deviate
dramatically from the recipes assumed in \citet{DeLucia07}, where
$\tau_2$ is a function of the cold gas mass and metallicity.  We will
be most interested in differential effects, so details of the
dependence of $\tau_2$ on galaxy properties is less of a concern here
than elsewhere.

\vspace{0.5cm}

\section{Results: Uncertainties in Translating Synthetic Galaxies into
  observables}
\label{s:res}

This section demonstrates the difficulty of translating the synthetic
galaxies produced from galaxy evolution models into observables.
Discussion is focused around three colors, one of each sensitive to
the ultraviolet, optical, and near--infrared regions of the spectrum.
We will first investigate the dependence of these colors on the
parameters in our dust model, and then the parameters controlling the
SSPs.  We will then combine all uncertainties to demonstrate their net
effect on the derived colors, and will explore the effect of these
uncertainties on model LFs.  Discussion is focused on broad--band
filters available from the {\it GALEX} \citep{Martin05}, SDSS
\citep{York00}, and 2MASS \citep{Jarrett00} surveys.  For reference,
the filters considered and their effective wavelengths are
$NUV=2300$\AA, $g=4700$\AA, $r=6200$\AA, and $K=2.2\mu m$.

\vspace{0.5cm}

\subsection{Dependence of synthetic galaxy colors on uncertain
  SPS ingredients}
\label{s:res1}

\subsubsection{Dust uncertainties}

Figure \ref{fig:dust} shows the colors of two synthetic galaxies as a
function of the parameters of our dust model.  For each panel, only
one parameter is varied.  The fiducial set of parameters is:
$\delta=0.7$, $\tau_1=1.0$, $\tau_2=0.3$, $t_{\rm esc}=7.0$,
$\sigma_{\rm clump}=0.0$, $\Delta_L=0.0$, $\Delta_T=0.0$, $f_{\rm
  BHB}=0.0$, $S_{\rm BS}=0.0$, and $\alpha_{\rm IMF}=2.3$.  Several
important trends are apparent.  First, the effect of $\tau_2$ on the
colors is the strongest of the parameters explored.  This is not
surprising as $\tau_2$ controls the total amount of attenuation in the
diffuse component.  The slope of the attenuation curve, $\delta$, and
the timescale for escape from the birth cloud, $t_{\rm esc}$, are
similar in that they have a substantial effect in the UV, with
increasingly diminishing importance toward the near--IR.  The
parameters $\tau_1$ and $t_{\rm esc}$, which control the attenuation
around young stars, have no effect on the passive galaxy because that
galaxy has no young stars.  The parameter $\tau_1$ saturates at high
values for the star--forming galaxy because, as one completely
extinguishes the light from young systems (where $t\leq t_{\rm esc}$),
the colors will asymptote to the colors of the older population, which
in this case are still relatively blue.

Finally, it is clear that the clumpiness of the ISM, controlled by
$\sigma_{\rm clump}$, has the effect of making colors bluer, and its
effect tends to saturate at $\sigma_{\rm clump}\sim1.3$.  Colors
become bluer as $\sigma_{\rm clump}$ increases because the average
optical depth must decrease for a clumpier medium (see Equation
\ref{eqn:sigmaclump}).  The effect saturates when the average optical
depth approaches zero, which occurs at $\sigma_{\rm clump}\sim1.3$
(again see Equation \ref{eqn:sigmaclump}).  In this case most stars
are unobscured and therefore a further increase in $\sigma_{\rm
  clump}$ cannot alter the integrated colors of the whole population.

The importance of the clumpiness of the ISM is explored further in
Figure \ref{fig:clumps}.  This figure is similar to the far--right
column of Figure \ref{fig:dust}, except here we show results only for
the star--forming galaxy, for varying values of $\tau_2$ (notice also
that the range of the y--axes are different).  It is clear that modest
increases in $\tau_2$ result in a substantially stronger dependence of
colors on the clumpiness of the ISM.  This trend is due to the fact
that increasing $\sigma_{\rm clump}$ decreases the effective dust
opacity to approximately it's limiting value of zero (as can be seen
by comparing columns three and five in Figure \ref{fig:dust}).  Since
variation of $\sigma_{\rm clump}$ can be interpreted as a variation of
the effective dust opacity from $\tau_2$ to $\approx0.0$, increasing
$\tau_2$ will clearly increase the effect of $\sigma_{\rm clump}$ on
the derived colors.

\begin{figure}[!t]
\plotone{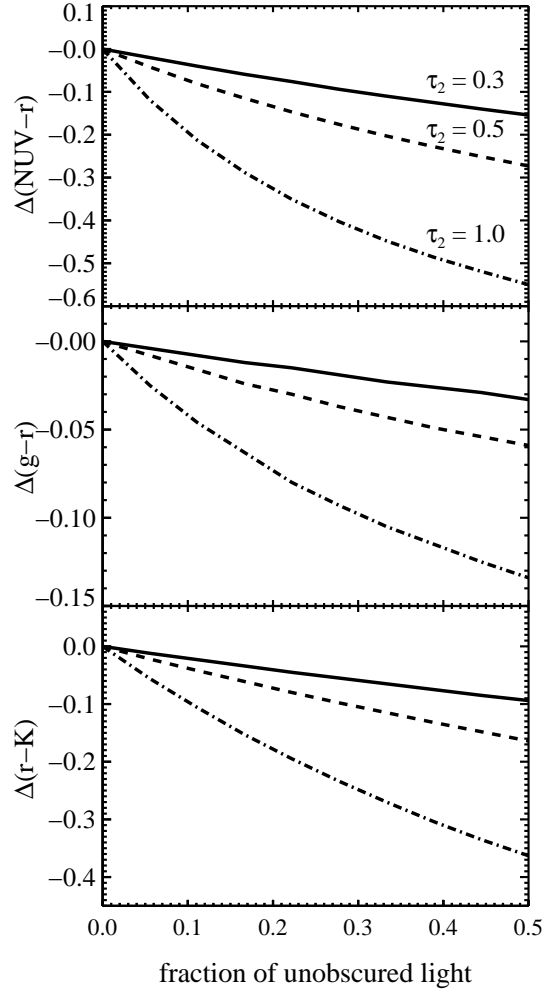}
\vspace{0.5cm}
\caption{Dependence of colors on the fraction of starlight unobscured
  by dust, $f$ (a fraction of $0.0$ corresponds to our standard model
  where all starlight is subject to attenuation).  Each relation is
  normalized to the colors at $f=0.0$.  The relations are shown for
  three values of the optical depth parameter $\tau_2$, as labeled in
  the figure. These results are shown for the star--forming galaxy
  only.}
\label{fig:fracd}
\end{figure}

\begin{figure*}[!t]
\center
\includegraphics[angle=90, width=0.9\textwidth]{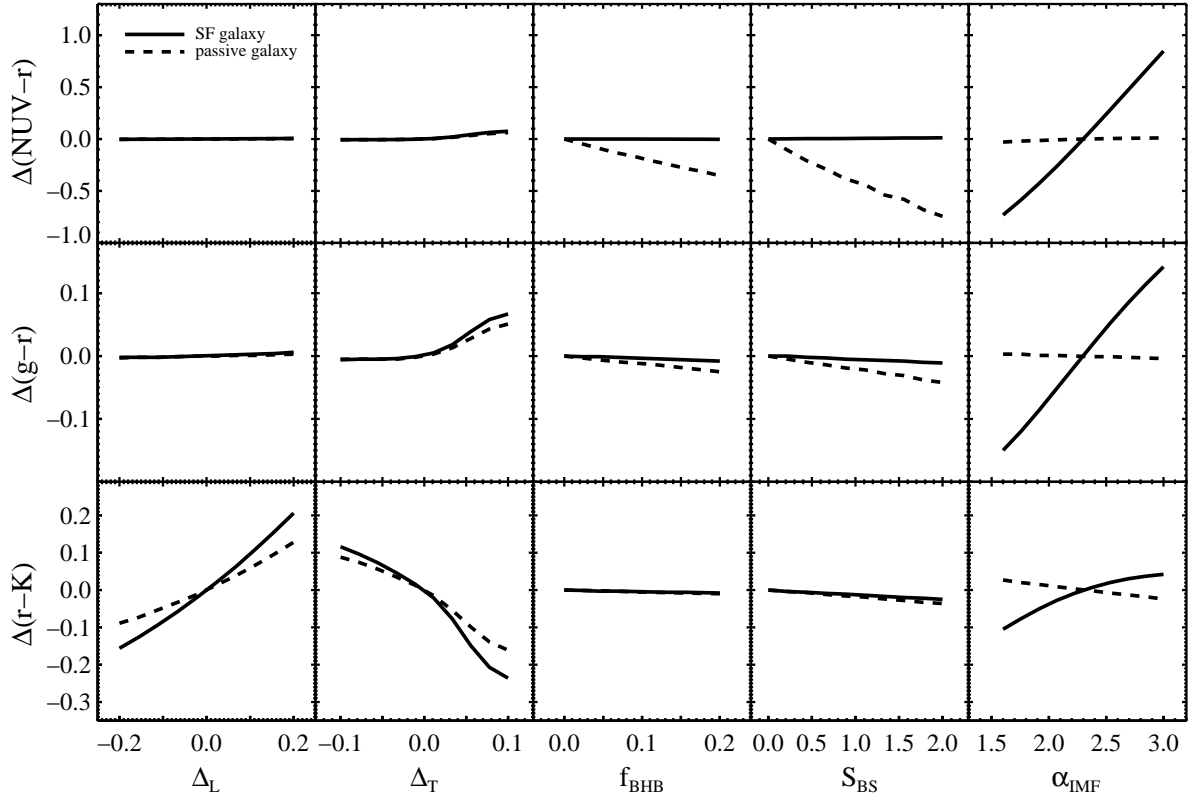}
\vspace{0.5cm}
\caption{Effect of SSP parameters on UV, optical, and near--IR colors
  of a star--forming galaxy ({\it solid lines}) and a passive galaxy
  ({\it dashed lines}).  The parameters are described in
  $\S$\ref{s:prior}.  The colors are shown as differences from a
  default model with $\Delta_L=0.0$, $\Delta_T=0.0$, $f_{\rm
    BHB}=0.0$, $S_{\rm BS}=0.0$, $\alpha_{\rm IMF}=2.3$, and with the
  default dust parameters as in Figure \ref{fig:dust}.  The population
  of blue horizontal branch stars and blue straggler stars, controlled
  by the parameters $f_{BHB}$ and $S_{BS}$, have little effect on the
  colors of the star--forming galaxy because this galaxy contains a
  large number of young hot stars that outshine these hot, evolved
  populations.  The IMF slope, $\alpha_{\rm IMF}$, is only varied for
  masses $>1\Msun$.}
\label{fig:isoc}
\vspace{0.5cm}
\end{figure*}

The implications of Figure \ref{fig:clumps} should not be
underestimated.  For galaxies that are moderately dusty
($\tau_2\gtrsim1$), the geometry of the dust is enormously important
when translating physical properties into observations, and visa
versa.  Note also that the dependence of colors on $\sigma_{\rm
  clump}$ is strongest for modest deviations from a uniform
distribution of dust.  Observations suggest that in the Milky Way
$\sigma_{\rm clump}\sim0.3$ \citep{Berkhuijsen08}, where the
dependence of color on $\sigma_{\rm clump}$ is strongest.  This
suggests that the detailed distribution of dust with respect to the
stars must be known to high precision.  Given this sensitivity, it is
reasonable to suspect that deviations from our assumption of a
lognormal distribution of column densities will also be important.

In Figure \ref{fig:fracd} we explore further the importance of the
large--scale spatial distribution of the dust with respect to the
stars.  In this figure we modify our default dust model by allowing
for a fraction of starlight to be unobscured by diffuse dust (the
aspect of the dust model regarding young stars is unchanged).  As
discussed in $\S$\ref{s:clumps}, realistic galaxies will likely have a
substantial fraction of starlight unobscured by dust.  Consider again
the example of a galaxy composed of a disk and a bulge with a majority
of its cold gas and dust at the midplane of the disk.  At any viewing
angle, roughly one half of the disk stars, and the majority of the
bulge stars, will be unobscured by dust.  Unobscured light fractions
as high as 50\% will therefore be common in galaxies containing dust.
Figure \ref{fig:fracd} further demonstrates that fruitfully comparing
galaxy models to observations requires a detailed understanding of the
relation between dust and stars.

\vspace{0.5cm}

\subsubsection{Stellar evolution and IMF uncertainties}
\label{s:stevimf}

\begin{figure*}[!t]
\plottwo{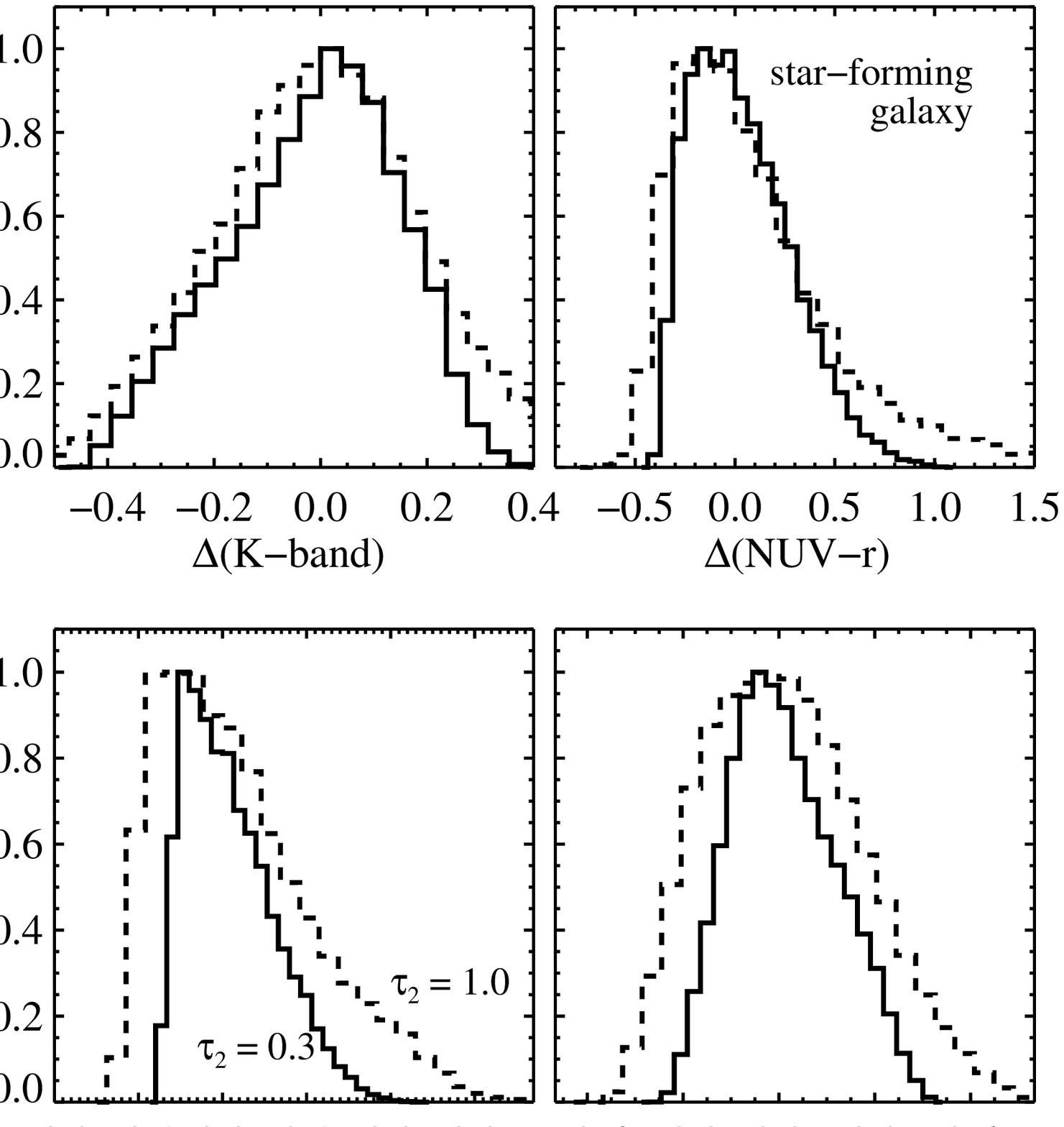}{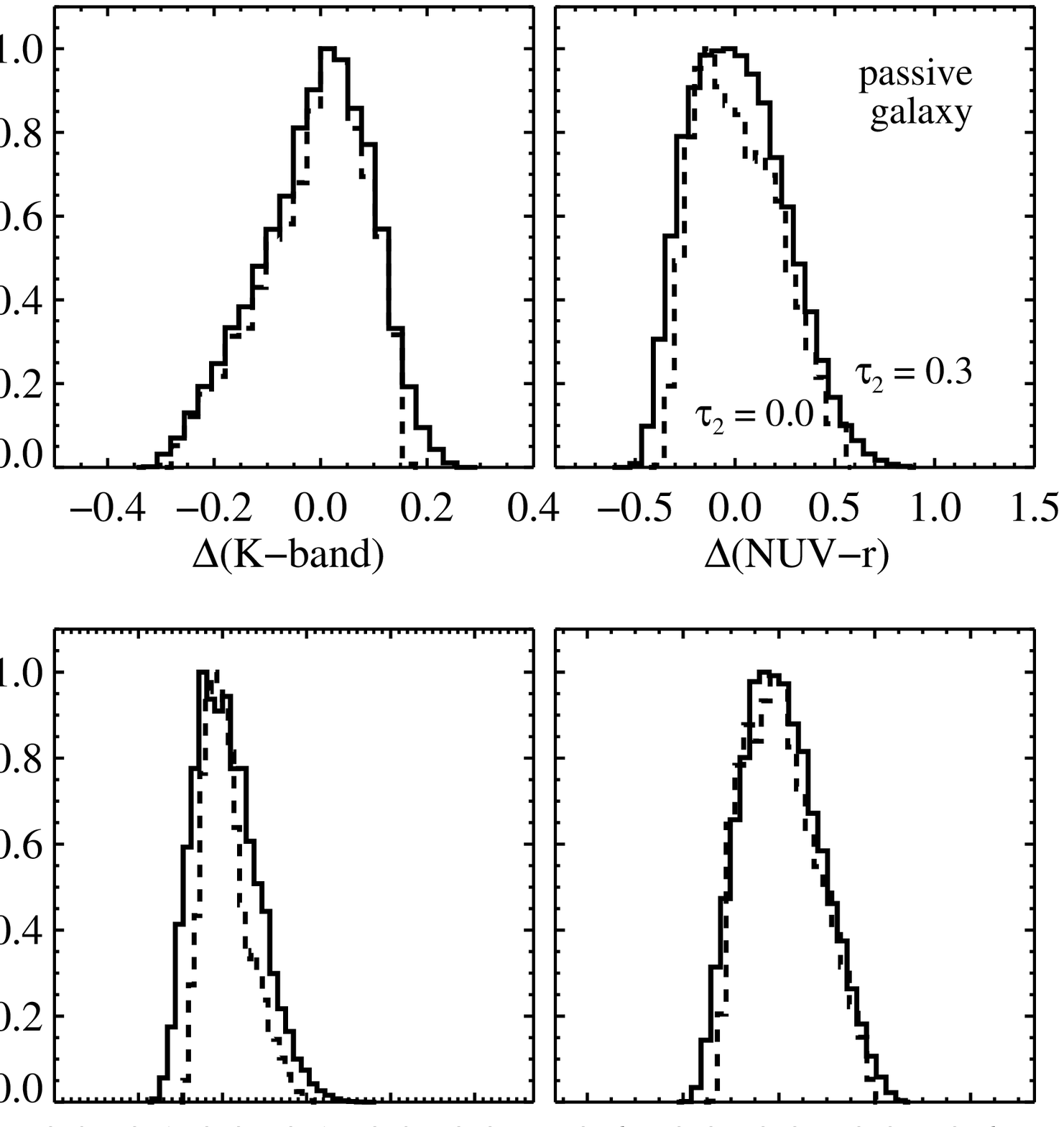}
\vspace{0.5cm}
\caption{Distribution of $K-$band magnitudes and broad--band colors
  arising from marginalizing over uncertainties in the dust model and
  stellar evolution.  All quantities are plotted as differences from
  our default model.  These distributions are created by marginalizing
  over all parameters shown in Figures \ref{fig:dust} and
  \ref{fig:isoc}, except for $\alpha_{\rm IMF}$ and $\tau_2$, for
  reasons discussed in the text.  {\it Left Panel:} Distributions for
  a star--forming galaxy.  Results are shown for two values of
  $\tau_2$: $\tau_2=0.3$ ({\it solid lines}), and $\tau_2=1.0$ ({\it
    dashed lines}). The difference between the two lines is driven by
  the clumpiness of the ISM, and is explained by the results in Figure
  \ref{fig:clumps}.  {\it Right Panel:} Uncertainties for a passive
  galaxy.  Results are shown for two values of $\tau_2$: $\tau_2=0.3$
  ({\it solid lines}), and $\tau_2=0.0$ ({\it dashed lines}).  Note
  that in the latter case this implies that the galaxy is entirely
  dust--free.}
\label{fig:tot}
\vspace{0.5cm}
\end{figure*}

Figure \ref{fig:isoc} shows the colors of two synthetic galaxies as a
function of the parameters controlling the SSPs, i.e. parameters
governing late stages of stellar evolution and the logarithmic slope
of the IMF at high masses ($>1\Msun$).  It is clear that the TP--AGB
parameters, $\Delta_L$ and $\Delta_T$, have little or no impact in the
UV and optical, but a large effect in the near--IR.  The effect is
larger for the star--forming galaxy because this galaxy has a larger
fraction of intermediate age stars, and the TP--AGB phase contributes
substantially to the bolometric luminosity of intermediate age
populations \citep{Maraston05, Conroy09a}.

The morphology of the horizontal branch and the frequency of blue
straggler stars, parameterized by $f_{\rm BHB}$ and $S_{\rm BS}$,
respectively, only impact the colors of the passive galaxy.  These
parameters have no effect on the star--forming galaxy because this
galaxy has young, hot stars that outshine the hot, but less luminous,
blue straggler and BHB stars.  The impact of these parameters on the
UV light for the passive galaxy is quite strong.  Even a modest number
of blue horizontal branch or blue straggler stars can produce changes
in the UV of several tenths of a magnitude.  The impact of these
exotic stars on the UV spectrum of passive galaxies has been discussed
for decades as a possible explanation for the UV upturn phenomenon in
elliptical galaxies \citep{Burstein88, Dorman95, Han07}.

The final column of Figure \ref{fig:isoc} shows the effect of the
logarithmic slope of the IMF on the derived colors of synthetic
galaxies.  The slope is only allowed to vary for stellar masses
$>1\Msun$.  The explanation of the trends seen in the figure are
qualitatively different for the two galaxy types.  The trends seen for
the passive galaxy are explained by inspection of the IMF--dependence
of the SSPs \citep[see][for details]{Conroy09a}.  For wavelengths
$\lesssim5000$\AA, the SED is sensitive to stars at a particular mass
(i.e. stars near the main--sequence turn--off mass).  However, at
wavelengths $\gtrsim5000$\AA, the SED is sensitive to both the red
giant branch (RGB) and asymptotic giant branches (AGBs), and in
particular the TP--AGB phase.  For a coeval set of stars, these two
phases of stellar evolution are inhabited by stars of slightly
different masses (stars along the AGB are more massive than stars
along the RGB), and therefore the relative weights given to these
phases will depend on the IMF.  In essence, a steeper IMF favors the
RGB over the AGB, and since the AGB is at lower effective temperature,
a steeper IMF results in slightly bluer near--IR colors.

The trends of $\alpha_{\rm IMF}$ with color for the star--forming
galaxy are both much stronger and, in the optical and near--IR, in the
opposite sense compared to the passive galaxy.  These trends are due
to the fact that the colors of a star--forming galaxy are sensitive to
stars of a range of masses.  This can be understood from the following
example.  Imagine that a galaxy consists of two SSPs: an old and a
young population.  As mentioned above, the light from each component
is dominated by stars of a given (different) mass, and so the relative
importance of each young and old component is determined by both the
star formation history, which specifies the fraction of mass formed in
each component, and the IMF, which specifies the relative weights
given to the mass intervals that dominate each component.  One can see
this another way by considering an analytic representation of the
combined flux:
\noindent
\be
F_\lambda \sim \Phi(M_1)\Psi(t_1)S_\lambda(M_1,t_1) + \Phi(M_2)\Psi(t_2)S_\lambda(M_2,t_2),
\ee
\noindent
which is qualitatively analogous to Equation \ref{eqn:flux}, without
dust.  This equation is approximate insofar as we are assuming that
the old population, born at time $t_1$, is dominated by stars of mass
$M_1$, while the young population born at time $t_2$ is dominated by
stars of mass $M_2$.  The logarithmic slope of the IMF determines the
ratio $\Phi(M_2)/\Phi(M_1)$, and it is therefore clear that the IMF
directly effects the emergent flux for a fixed star formation history.

In the context of the above example, increasing $\alpha_{\rm IMF}$
results in a redder SED because the young (blue) component is
sensitive to high--mass stars, while the old (red) component is
sensitive to low--mass stars, and a steeper IMF favors low--mass over
high--mass stars.  This simple example serves to qualitatively explain
the trends observed in the far--right column of Figure \ref{fig:isoc}.

\vspace{0.5cm}

\subsection{Total uncertainties in UV, optical, and near--IR colors}

In the previous section we explored the dependence of synthetic galaxy
colors on various uncertain aspects of the SPS technique, including
uncertainties in the dust model, stellar evolution, and the IMF.  We
are now in a position to investigate the combined effects of these
uncertainties on the broad--band colors of synthetic galaxies.

Figure \ref{fig:tot} shows the distribution of colors and $K-$band
magnitudes for two synthetic galaxies.  The distributions were
obtained by marginalizing over the uncertain aspects of SPS discussed
in the previous section.  All quantities are plotted as differences
with respect to our default model.  We marginalize over all parameters
shown in Figures \ref{fig:dust} and \ref{fig:isoc} {\it except} for
$\alpha_{\rm IMF}$ and $\tau_2$.  We choose to fix $\tau_2$ because,
if we interpret this parameter as probing the total dust mass, then we
may hope that galaxy evolution models are able to predict this
quantity, if not at present then in the future.  In other words, our
aim in this work is to marginalize over aspects of SPS that must be
{\it assumed} when translating models into observations.  If one or
more parameters discussed in the previous section is readily {\it
  predicted} by such models, then such parameters should obviously not
be marginalized over.

We choose to fix $\alpha_{\rm IMF}$ simply because its effect on the
resulting colors is so large that it would dominate the error budget
in all cases.  The uncertainties in broad--band colors inferred from
Figure \ref{fig:tot} can thus be safely interpreted as lower limits to
the true uncertainty.

Figure \ref{fig:tot} shows results for different values of $\tau_2$.
For both galaxies we consider our default value of $\tau_2=0.3$.  For
the star--forming galaxy we also consider $\tau_2=1.0$, while for the
passive galaxy we consider $\tau_2=0.0$.  We vary $\tau_2$ for the
star--forming galaxy to demonstrate the fact that the relationship
between $\sigma_{\rm clump}$ and the resulting colors itself depends
on $\tau_2$ (cf. Figure \ref{fig:clumps}).  Larger values of $\tau_2$
imply that variation of $\sigma_{\rm clump}$ will have a progressively
larger impact on the derived colors, as is evident in the broader
distributions in Figure \ref{fig:tot} for $\tau_2=1.0$.

Passive galaxies are often observed to contain little dust
\citep[$\tau_2\lesssim0.1$; see e.g.,][]{Goudfrooij94, Ferrari99,
  Temi04, Draine07}.  We thus also include $\tau_2=0.0$ in Figure
\ref{fig:tot} for the passive galaxy, which implies that the
parameters $\sigma_{\rm clump}$ and $\delta$ have no effect in the
marginalization.  The removal of dust decreases the uncertainties only
slightly.  This is because, for the passive galaxy, uncertainties are
dominated by the uncertainties in stellar evolution.

\begin{figure*}[!t]
\plotone{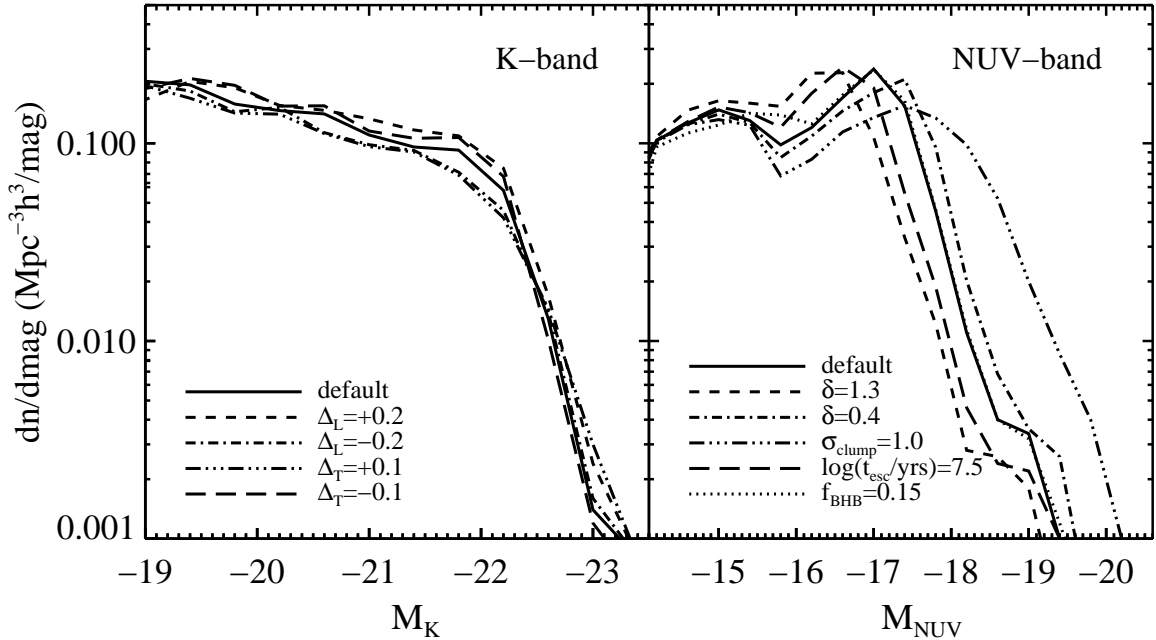}
\vspace{0.5cm}
\caption{LFs of synthetic galaxies from the galaxy formation model of
  \citet{DeLucia07}.  SFHs and metallicity histories are provided by
  the model.  For the same set of galaxies, LFs are constructed in the
  $K-$band and $NUV-$band for a variety of assumptions regarding the
  dust model (parameterized by $\delta$, $\sigma_{\rm clump}$, and
  $t_{\rm esc}$), importance of the TP--AGB stellar evolutionary phase
  ($\Delta_L$ and $\Delta_T$), and fraction of blue HB stars ($f_{\rm
    BHB}$).  It is clear that for the same galaxy formation model, the
  predicted LFs can span a wide range depending on the details of how
  one translates the model galaxies into the observational plane.}
\label{fig:mrlf}
\vspace{0.5cm}
\end{figure*}

LFs constructed from a subsample of the full galaxy formation model
are shown in Figure \ref{fig:mrlf}, both in the $K-$band and
$NUV-$band.  The LFs are computed from the model star formation and
metallicity histories, using the SPS model described in
$\S$\ref{s:sps}.  LFs are constructed for a variety of assumptions
regarding the stellar evolution and dust model parameters.  Note that
the LFs were constructed from the exact same galaxy sample in each
case, so the variations seen in Figure \ref{fig:mrlf} are entirely due
to the different modeling assumptions (as opposed to Poisson or sample
variance).

Clearly, the TP--AGB parameters $\Delta_L$ and $\Delta_T$ have a large
and non--trivial effect on the $K-$band LF.  The dust parameters have
a negligible effect on the $K-$band LF, and so their variation is not
included in the $K-$band LF.  In contrast, the dust model parameters
have a dramatic effect on the $NUV-$band LF.  The TP--AGB parameters
have no effect in the $NUV$.  While not shown, sensitivity to the
ensemble of parameters is roughly minimized in the $r-$band,
suggesting that the optical portion of the SED is the most robust to
SPS uncertainties.

Notice that for both LFs in Figure \ref{fig:mrlf} the {\it shape} of
the LF changes for different model parameters.  This is due to the
fact that the effects of these parameters vary with stellar population
age.  The effects of SPS uncertainties therefore induce complex
systematic uncertainties in model LFs in the UV and near-IR that must
be carefully handled if robust constraints on galaxy formation models
are to be achieved.

\section{Discussion}
\label{s:disc}

\subsection{How do we interpret these uncertainties?}

We have demonstrated that the broad--band colors of synthetic galaxies
carry significant uncertainties due to uncertain aspects of SPS.
These uncertainties must thus be incorporated into any analysis that
attempts to compare models to observations.  But how are these
uncertainties to be interpreted?

Unfortunately, it is not clear whether or not one should treat these
uncertainties as statistical or systematic.  By statistical we mean
that the uncertain SPS parameter values vary stochastically from galaxy
to galaxy, whereas by systematic we mean that these parameters vary
systematically across galaxy types.  Uncertainties associated with the
dust model, including the attenuation curve and large--scale geometry
of the ISM, can probably be interpreted as statistical uncertainties.
These aspects depend in very complex and uncertain ways on the
geometry of the stars with respect to the dust, the local population
of O--type stars, the gas--phase metallicity, and other physical
properties of the ISM.

Uncertainties associated with stellar evolution are more likely
systematic, although there may be a statistical component.  One might
imagine that the fraction of blue straggler or horizontal branch
stars, or the evolution of TP-AGB stars, would not vary stochastically
from galaxy to galaxy since stellar evolution is believed to depend on
a small number of variables.  However, even within globular clusters
there appears to be a need for stochastic mass--loss along the RGB in
order to explain the observed broad range in temperatures along the
horizontal branch \citep[e.g.,][]{Dorman95, Kalirai07}.  Thus, in
addition to the systematic variation of these post--main sequence
stars with global quantities such as metallicity and age, one might
indeed expect a statistical uncertainty as well.

Uncertainties in the IMF may be plausibly be either statistical or
systematic, depending on the dominant mechanisms responsible for the
shape of the IMF (i.e., is gas--phase metallicity or ISM pressure more
important?).

Whether or not the resulting color uncertainties are predominantly
statistical or systematic obviously has very different implications
for their interpretation.  For the statistical errors, the
interpretation is straightforward.  Systematic uncertainties are
rather more troubling in this context because systematics will be
correlated with galaxy properties such as age, metallicity, and/or
mass.  For example, if future observations of TP--AGB stars imply that
$\Delta_L=0.3$, then the colors will shift in an age--dependent (and
because mass correlates with age, a mass--dependent) way.  Thus,
systematic uncertainties will tend to introduce not only overall
shifts but also tilts in observational quantities derived from models
such as the color-magnitude diagram and LFs.

With the preceding discussion in mind, we return to Figure
\ref{fig:tot}.  It should be clear that the distributions shown in
this figure are {\it not} to be interpreted as PDFs for colors given a
star formation history.  Consider a simple example.  If we had chosen
to marginalize over nine parameters that had almost no effect on a
particular color and one parameter that had a large effect, then the
distribution in colors would be very strongly peaked at
$\Delta($color$)\approx0$, with broad wings at low frequency, because
each parameter is sampled equally.  Even with a strongly peaked
distribution, one would not conclude that the uncertainty on the color
is small, since that one important parameter produces such large
uncertainties.  The implication is that the only robust way to
interpret these distributions is to consider their {\it entire width}
as a simple measure of the resolution with which one may accurately
predict colors given star formation and metal enrichment histories.

\subsection{Why the IMF matters}

It is often assumed that results are insensitive to the IMF so long as
both modelers and observers use the same IMF to translate between
fluxes and physical parameters.  This assumption is incorrect.  Galaxy
evolution models predict star formation histories (SFHs) that are
independent of the IMF.  Upon integration, these SFHs provide total
stellar masses for model galaxies.  In this context, the IMF enters
only in the estimation of the fraction of mass returned to the ISM via
winds and supernovae.

In contrast, since the IMF has a direct effect on the colors of
star--forming galaxies, it will effect the SFRs and stellar masses
inferred from observations, and will effect modelers' ability to
translate and compare their models to observations.  It is important
to realize that this is true even if the low-mass end of the IMF ---
which has a negligible effect on the integrated light from galaxies
--- is fixed.  It is the slope of the IMF at masses $>1\Msun$ that
most affects the colors of star--forming galaxies.

The IMF is thus extraordinarily important when attempting to compare
models to observations.  As we have demonstrated herein, the measured
uncertainty in the IMF in the solar neighborhood is sufficiently large
to adversely effect comparisons between models and observations.  A
more pessimistic assessment of our knowledge of the IMF in external
galaxies (and at earlier epochs), will only exacerbate this problem.

\subsection{What is to be done?}

Despite the magnitude of the uncertainties presented herein, there is,
we believe, a clear way to proceed in the comparison between models
and observations.

As emphasized in the present work, the principle difficulty in
translating models into observables is the array of important aspects
of SPS that are neither well--constrained observationally nor
well--understood theoretically.  We are thus forced to adopt rough
ranges for these uncertainties --- rough because uncertainties are
difficult to quantify.  However, this does not imply that each {\it
  observed} galaxy can be equally well--fit by the full range of
parameters.  In other words, it is precisely because these parameters
impact the broad--band colors of galaxies that they may be
constrained, to some extent, directly by the data.

We therefore advocate using the SPS technique to estimate the basic
physical properties of observed galaxies, including star formation
rates and stellar masses.  The SPS technique must include a
marginalization over the uncertain aspects discussed herein.  The
derived properties will therefore be robust, though they will carry
larger errors.  Such an approach requires a pan--chromatic view of a
large number of galaxies at multiple epochs.  The physical properties
of observed galaxies can then be directly compared to galaxy evolution
models, so long as proper attention is given to the impact of the
associated uncertainties.  We believe that this approach will afford a
more accurate and stringent constraint on models than the approach of
converting models into observables.

This recommendation is guided largely by the manner in which modelers,
in practice, compare their predictions to observations.  Models are
often tested against observed LFs, color--magnitude diagrams, and
luminosity--dependent clustering.  We instead suggest that models
should be tested against observationally--constrained physical
relations including mass functions, mass--dependent clustering, and
SFR--mass correlations.  As these observational results have become
available in the past few years, modelers have begun to use them as
constraints.  We believe that this is the better way to proceed, as
long as two conditions are met: 1) the observational results contain
an accurate accounting of all the relevant uncertainties, including
the correlations between derived products, and 2) the comparison
between model and observations takes careful account of the resulting
uncertainties on the derived physical properties.

Proponents of the alternative approach, namely that modelers translate
their results into the observational plane, may counter that in the
models the SFH and metallicity of the stars is known precisely,
whereas in the data these aspects must be estimated from the data.
The age--metallicity degeneracy \citep{Worthey94}, for example, may
present an ambiguity in the fitted data, whereas none would arise by
converting the models into the observational plane.  We believe that
the fitting the SFH and metallicity in the data, in addition to
fitting for dust and SPS uncertainties, is still the preferred
approach, rather than marginalizing over the uncertain dust model
parameters and SPS uncertainties when attempting to robustly translate
model results into observational predictions.  This view is supported
by the recent work of \citet{Gallazzi09} who find that in the absence
of dust, uncertainties in fitting SFHs and metallicities translate
into a relatively minor uncertainty in the stellar mass--to--light
ratio.  These authors conclude that dust obscuration and SPS
uncertainties remain the dominant uncertainties when estimating
physical parameters of galaxies.

\section{Summary}
\label{s:sum}

We have investigated some of the uncertainties associated with
translating synthetic galaxies into observables, focusing on
broad--band UV through near-IR photometry available from the {\it
  GALEX}, SDSS, and 2MASS surveys.  The uncertainties associated with
stellar evolution and dust are important and effect passive and
star--forming galaxies in different ways.

For star--forming galaxies, both uncertainties in dust and in the
TP--AGB phase of stellar evolution result in substantial uncertainties
in UV, optical, and near--IR colors.  Uncertainties in the spatial
distribution of dust with respect to stars play an especially
important role, as do uncertainties in the dust attenuation law.

The uncertainties in the colors of passive galaxies are dominated by
uncertainties in stellar evolution, including the TP--AGB phase, blue
stragglers, and the morphology of the horizontal branch.  The
uncertainties in the dust model have a minor effect because the total
dust content of passive galaxies is observed to be low, though
non--zero \citep[e.g.][]{Goudfrooij94, Ferrari99, Draine07}.

We have also explored the impact of the IMF on the derived colors of
galaxies, although we have not included this source of uncertainty in
the total error budget.  Varying the logarithmic slope of the IMF has
different effects for passive and star--forming galaxies.  The
magnitude of the effect is much larger for star--forming galaxies, and
needs to be taken seriously by both modelers and observers.  The
importance of the IMF in this regard has been known since at least the
work of \citet{Tinsley80} --- we have explored it here primarily to
(re)call attention to its importance when comparing models to
observations.

The ranges of uncertainties investigated here are largely optimistic
(i.e. they are relatively small), and yet their influence on our
ability to fruitfully compare models to observations is substantial.
Many of these uncertainties are systematic, further complicating their
interpretation.  The comparison of models to observations can be done
with confidence only once these uncertainties are carefully included
in the analysis.  The results from this work highlight the urgent need
for observations capable of diminishing the various uncertainties
discussed herein.

\acknowledgments 

We thank the Virgo Consortium for making the Millennium Simulation and
related semi-analytic galaxy formation model publicly available, Bruce
Draine for stimulating discussions regarding dust properties, and
Guinevere Kauffmann for comments on an earlier draft.  The referee is
thanked for a thoughtful report that clarified several topics.  This
work made extensive use of the NASA Astrophysics Data System and of
the {\tt astro-ph} preprint archive at {\tt arXiv.org}.


\end{document}